\documentclass[floatfix,twocolumn,showpacs,floats,aps,prl,showpacs,amsfonts,amssymb]{revtex4}
\usepackage{amssymb}
\usepackage{bm}
\usepackage{hyphenat}
\usepackage{graphicx}
\usepackage{dcolumn}
\newlength{\sepmod}
\def\eqref#1{(\ref{#1})}

\def\d{\mathrm{d}}

\begin{document}
\title{Cross-correlations in scaling analyses of phase transitions}

\author{Martin Weigel}
\email{weigel@uni-mainz.de}
\affiliation{Institut f\"ur Physik, Johannes Gutenberg-Universit\"at Mainz,
  Staudinger Weg 7, D-55099 Mainz, Germany}

\author{Wolfhard Janke}
\email{janke@itp.uni-leipzig.de}
\affiliation{Institut f\"ur Theoretische Physik and Centre for Theoretical Sciences (NTZ),
  Universit\"at Leipzig, Postfach 100\,920, D-04009 Leipzig, Germany}

\date{\today}

\begin{abstract}
  Thermal or finite-size scaling analyses of importance sampling Monte Carlo time
  series in the vicinity of phase transition points often combine different estimates
  for the same quantity, such as a critical exponent, with the intent to reduce
  statistical fluctuations. We point out that the origin of such estimates in the
  same time series results in often pronounced cross-correlations which are usually
  ignored even in high-precision studies, generically leading to significant
  underestimation of statistical fluctuations. We suggest to use a simple extension of the
  conventional analysis taking correlation effects into account, which leads to
  improved estimators with often substantially reduced statistical fluctuations at
  almost no extra cost in terms of computation time.
\end{abstract}

%05.10.Ln  Monte Carlo methods
%75.50.Lk Spin glasses and other random magnets
%64.60.Fr Equilibrium properties near critical points, critical exponents
%05.10.Cc Renormalization group methods
%02.60.Pn Numerical optimization
%75.10.Hk Classical spin models
%05.70.Fh  Phase transitions: general studies 

\maketitle

With the advent of the renormalization group in the 1960s, the notions of scaling and
universality have been combined into the solid basis of our understanding of critical
phenomena in statistical physics, field theory and a wealth of applications in areas
ranging from solid-state physics \cite{chaikin:00} to cosmology \cite{bergstrom:04}.
It is through the remarkable fact that the most important properties of a continuous
phase transition are independent of many microscopic details and, instead, only
depend on a small number of fundamental characteristics of a system, such as the
dimensionality and the symmetries of the order parameter, that we can accurately
describe such different situations as, e.g., the liquid-vapor phase transition and
the ferromagnetic transition of uniaxial magnets, with one and the same scaling
theory. Recently, the investigation of quantum phase transitions has opened up a new
Pandora's box with a wealth of transitions partially defining novel universality
classes \cite{vojta:03b}. The direct applicability of results from simple models to a
range of experimentally realized systems implied by the principle of universality
renders high-precision determinations of critical parameters for the most common
universality classes a rewarding goal.

Particularly through the conception of advanced finite-size scaling (FSS) approaches
and novel efficient algorithms \cite{binder:book2}, Monte Carlo (MC) simulations have
grown up to become a tool for the determination of universal critical quantities
clearly competitive compared to the more traditional approaches of high-temperature
and field-theoretic expansions \cite{pelissetto:02}. Likewise, the detailed
investigation of systems undergoing first-order phase transitions has become a
classic application of the MC simulation technique \cite{janke:03}. Major advances in
the competitiveness of the MC method for these purposes came with the advent of
histogram methods \cite{ferrenberg} and generalized-ensemble simulation techniques
such as the multicanonical method \cite{berg:92b}, both of which allow for extracting
estimates of thermal averages for a continuous range of temperatures or other
external parameters from a {\em single\/} MC simulation. It is only through this
effective continuity of information that high-precision investigations of phase
transitions have come into the reach of simulation methods.  For arriving at
high-precision estimates, however, all possible sources of error must be put under
close scrutiny. This is often done to a high degree of sophistication concerning the
systematic errors resulting from corrections to scaling \cite{pelissetto:02} and the
statistical errors resulting from the stochastic nature of MC time series (including
their timewise autocorrelations for the case of the most commonly used Markov chain
MC techniques) \cite{berg:04,efron:book}. It has not been systematically discussed
previously, however, that the extraction of different estimates from a single time
series in thermal or FSS analyses must entail cross-correlations. As will be shown
below, neglecting their effect not only results in systematically wrong estimates of
statistical errors, but also fails to fully exploit the available time-series data to
yield the maximum statistical precision obtainable.

Although our considerations apply generally to all situations where a number of
different estimates from the same (set of) simulation(s) are combined to a final
result, for specificity we consider the FSS analysis of simulation
data in the vicinity of a critical point. For the purpose of illustration we choose
the technique outlined in Ref.~\cite{ferrenberg:91a}, but very similar
considerations apply to alternative approaches, see, e.g.,
Refs.~\cite{ballesteros:96,hasenbusch:99}. To be specific, we here use a
magnetic language and first consider the maxima of the derivative of the
magnetization cumulants $U_{2k} = 1-\langle |m|^{2k}\rangle/3\langle|m|^k\rangle^2$
for $k=1$, $2$, $\ldots$:
\begin{equation}
  \label{eq:binder_scaling}
  \left.\frac{\d U_{2k}}{\d\beta}\right|_\mathrm{max} = U_{k,0}L^{1/\nu}(1+U_{k,c}L^{-w}+\cdots),
\end{equation}
where $L$ denotes the linear size of the system and $\beta$ is the inverse
temperature. This relation allows for a precise determination of the correlation
length exponent $\nu$ without previous knowledge of the critical temperature. In many
cases at least some of the scaling corrections, such as an effective leading
correction with exponent $w$ as indicated in Eq.~(\ref{eq:binder_scaling}), need to
be taken into account to achieve the desired level of accuracy. An analogous relation
holds for the scaling of the maxima of the logarithmic derivative of magnetization
moments,
\begin{equation}
  \label{eq:logmagnderiv_scaling}
  \left.\frac{\d \ln \langle |m|^k\rangle}{\d\beta}\right|_\mathrm{max} =
  D_{k,0}L^{1/\nu}(1+D_{k,c}L^{-w}+\cdots).
\end{equation}
These scaling relations for determining $\nu$ only become useful as soon as the
maximum values $\left.(\d U_{2k}/\d\beta)\right|_\mathrm{max}$ and $\left.(\d \ln
  \langle |m|^k\rangle/\d\beta)\right|_\mathrm{max}$ can be computed to high accuracy
without the need for repeated simulations manually tracking their locations in
$\beta$. In case of a histogram or reweighting analysis of a single canonical
simulation, this is effected through the continuous family of estimates
\begin{equation}
  \label{eq:reweighting}
  \hat{A}(\beta) = \frac{\sum_i A_i e^{-(\beta-\beta_0)E_i}}{\sum_i e^{-(\beta-\beta_0)E_i}}
\end{equation}
for the thermal average $\langle A\rangle_\beta$ from a time series $\{A_i\}$ of
measurements resulting from an importance sampling simulation at inverse temperature
$\beta_0$. Conventional techniques of numerical analysis such as a golden section
search then allow for an efficient determination of the maxima of
Eqs.~(\ref{eq:binder_scaling}) and (\ref{eq:logmagnderiv_scaling}) to high
precision.
%In some situations it may be impractical to store the whole time series
%$\{A_i\}$, and the exact reweighting relation (\ref{eq:reweighting}) might then be
%replaced by a Taylor expansion of $\langle A\rangle$ with respect to $\beta$, where
%the moments of $A$ appear as coefficients.
Once $\nu$ has been determined, the scaling of the shifts of the location of the
maxima of quantities such as $\d U_{2k}/\d\beta$ and $\d \ln \langle
|m|^k\rangle/\d\beta$ as well as the specific heat, susceptibility etc.\ allow to
locate the transition coupling $\beta_c$. Finally, the remaining critical exponents
may be estimated from the FSS of the maxima of the specific heat to yield
$\alpha/\nu$, of the susceptibility to yield $\gamma/\nu$ etc. Since the exponent
$\nu$ enters all FSS relations, it clearly is of utmost importance to exploit the
available data to their fullest for a precise estimate of $\nu$. In view of the
family of relations (\ref{eq:binder_scaling}) and (\ref{eq:logmagnderiv_scaling}),
this certainly includes a combination of estimates from $\d U_{2k}/\d\beta$ and $\d
\ln \langle |m|^k\rangle/\d\beta$ as well as from the different choices of the
parameter $k$ \cite{ferrenberg:91a}.

To see how this combination should be performed, consider a number $n$ of different
estimators $\hat{x}_i$ with the same expectation value (e.g., $\hat{x}_i
=\hat{\nu}_i$). A combined average results from a linear combination $\bar{x} =
\sum_i\alpha_i \hat{x}_i$ with $\sum_i \alpha_i = 1$. While any such combination
yields a valid estimator of $\langle x\rangle$, e.g., the arithmetic mean
$\bar{x}_\mathrm{plain}$ with $\alpha_i = 1/n$, the ensuing statistical fluctuations
will be larger than necessary. For {\em uncorrelated\/} estimates $\hat{x}_i$ minimal
variance of $\bar{x}$ is achieved for the error-weighted mean $\bar{x}_\mathrm{err}$
with \cite{brandt:book}
\begin{equation}
  \label{eq:error_weighted}
  \alpha_i = Z^{-1}\frac{1}{\sigma^2(\hat{x}_i)},
\end{equation}
where $\sigma^2(\hat{x}_i)$ denotes the variance of $\hat{x}_i$ and $Z = \sum_i
1/\sigma^2(\hat{x}_i)$. In general, however, the estimates $\hat{x}_i$, stemming from
a reweighting analysis of the same MC time series, will be substantially correlated.
Under these circumstances, the optimum choice is a covariance-weighted mean
$\bar{x}_\mathrm{cov}$ with weights \cite{brandt:book,janke:97}
\begin{equation}
  \label{eq:covariance_weighted}
  \alpha_i = Z^{-1} \sum_j [\Gamma(\hat{x})^{-1}]_{ji},
\end{equation}
where $\Gamma(\hat{x})^{-1}$ denotes the inverse of the covariance matrix
$\Gamma_{ij}(\hat{x}) = \langle \hat{x}_i
\hat{x}_j\rangle-\langle\hat{x}_i\rangle\langle \hat{x}_j\rangle$ and $Z = \sum_{ij}
[\Gamma(\hat{x})^{-1}]_{ij}$. Since for uncorrelated estimates
$[\Gamma(\hat{x})^{-1}]_{ij} = \delta_{ij}/\sigma^2(\hat{x}_i)$,
Eq.~\eqref{eq:error_weighted} is recovered in this special case. Even more
dramatically affected by correlations are the statistical errors of averages, where
the standard formula $\sigma^2_\mathrm{uncorr}(\bar{x}) = \sum_i \alpha_i^2
\sigma^2(\hat{x}_i)$ is no longer valid and must be modified to read
$\sigma^2_\mathrm{corr}(\bar{x}) = \sum_{i,j} \alpha_i\alpha_j \Gamma_{ij}(\hat{x})$,
generically leading to an underestimate of fluctuations via the naive (and wrong)
estimator $\sigma^2_\mathrm{uncorr}$.

To check for the strength of such correlation effects and their influence on finding
optimal averages endowed with valid estimates of statistical errors, we performed a
FSS analysis of the critical points of the ferromagnetic Ising model in two (2D) and
three (3D) dimensions. Time series data for the configurational energy and
magnetization were produced from one single-cluster update simulation
\cite{binder:book2} per system size at a fixed temperature. Estimates for the
exponent $\nu$ were extracted from FSS fits of the relations
(\ref{eq:binder_scaling}) resp.\ (\ref{eq:logmagnderiv_scaling}) to the maxima of $\d
U_{2k}/\d\beta$ with $k=1$ and $k=2$ as well as $\d \ln \langle |m|^k\rangle/\d\beta$
with $k=1$, $2$, and $3$ extracted from a reweighting analysis. Statistical errors
for the individual estimates were calculated via a jackknife analysis
\cite{efron:book} over the reweighting procedure, taking timewise autocorrelations
into account. Likewise, the covariance matrix $\Gamma$ was determined from the
non-parametric jackknife estimator known to be especially robust \cite{efron:book},
\begin{equation}
  \label{eq:COV}
  \widehat{\mathrm{COV}}(\hat{\nu}_i,\hat{\nu}_j) = \frac{n-1}{n}\sum_{s=1}^n
  [\hat{\nu}_{i(s)}-\hat{\nu}_{i(\cdot)}]
  [\hat{\nu}_{j(s)}-\hat{\nu}_{j(\cdot)}].
\end{equation}
Here, $n$ denotes the number of jackknife blocks, $\hat{\nu}_{i(s)}$ denotes the
value for jackknife block $s$ and $\hat{\nu}_{i(\cdot)}$ is the arithmetic average of
the $\hat{\nu}_{i(s)}$. For the results presented here, $n = 100$ blocks were used,
where we checked that the results are invariant, at the level of statistical
fluctuations, to the choice of a significantly larger number of blocks.

\begin{table}[tb]
  \centering
  \caption
  {
    Estimates of the correlation length exponent $\nu$ for the 2D and 3D Ising models
    from the scaling of the maxima (\ref{eq:binder_scaling}) and
    (\ref{eq:logmagnderiv_scaling}), as well as different averages and error estimates explained in
    the main text. 
%WJ The reference value for the 2D case is exact, whereas the value
%WJ for the 3D model is a meta-estimate taken from the recent review \cite{pelissetto:02}.
    The 2D reference value is exact, whereas in 3D
    it is taken from the recent review \cite{pelissetto:02}.
  }
  %\footnotesize
  \begin{ruledtabular}
    \begin{tabular}{lldddd}
      & & \multicolumn{2}{c}{2D} & \multicolumn{2}{c}{3D}\\
      & & \multicolumn{1}{c}{$\nu$} & \multicolumn{1}{c}{$\sigma$} &
      \multicolumn{1}{c}{$\nu$} &
      \multicolumn{1}{c}{$\sigma$} \\ \colrule
      \multicolumn{2}{c}{$\displaystyle \frac{\d\ln\langle |m|\rangle}{\d\beta}$} & 1.0085 & 0.0183 & 0.6358 & 0.0127 \\
      \multicolumn{2}{c}{$\displaystyle \frac{\d\ln\langle m^2\rangle}{\d\beta}$} & 1.0128 & 0.0194 & 0.6340 & 0.0086  \\
      \multicolumn{2}{c}{$\displaystyle \frac{\d\ln\langle |m|^3\rangle}{\d\beta}$} & 1.0175 & 0.0201 & 0.6326 & 0.0062 \\
      \multicolumn{2}{c}{$\displaystyle \frac{\d U_2}{\d\beta}$} & 1.0098 & 0.0281 & 0.6313 & 0.0020 \\
      \multicolumn{2}{c}{$\displaystyle \frac{\d U_4}{\d\beta}$} & 1.0149 & 0.0511 & 0.6330 & 0.0024\\ \colrule
      $\bar{x}_{\mathrm{plain}}$ & $\sigma_\mathrm{uncorr}$ & 1.0127 & 0.0141 & 0.6334 & 0.0038\\
      & $\sigma_\mathrm{corr}$ & & 0.0269 &  & 0.0067 \\
      $\bar{x}_{\mathrm{err}}$ & $\sigma_\mathrm{uncorr}$ & 1.0123 & 0.0102 & 0.6322 & 0.0015\\
      & $\sigma_\mathrm{corr}$ & & 0.0208 &  & 0.0024\\
      $\bar{x}_{\mathrm{cov}}$ & $\sigma_\mathrm{corr}$ & 0.9935 & 0.0078 & 0.6300 & 0.0017\\ \colrule
      \multicolumn{2}{c}{reference value} & 1 & & 0.6301 & 0.0004 \\
    \end{tabular}
  \end{ruledtabular}
  \label{tab:2D_3D_av}
\end{table}

\begin{figure}[t]
  \centering
  \includegraphics[keepaspectratio=true,scale=0.75,trim=45 48 75 78]{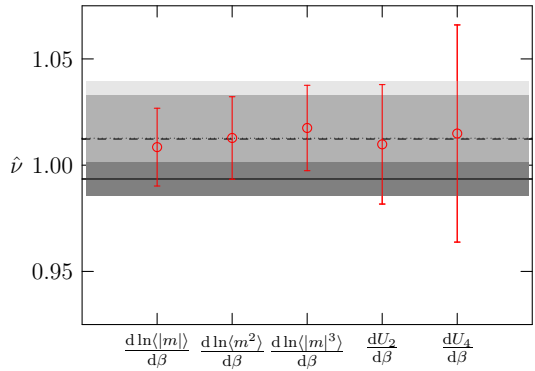}
  \caption
  {(Color online) Estimates of the critical exponent $\nu$ of the 2D Ising model from
    the FSS of the indicated observables. 
%WJ    The dotted line indicates
%WJ    the plain average $\bar{\nu}_\mathrm{plain}$, the dashed line stands for the
%WJ    error-weighted mean $\bar{\nu}_\mathrm{err}$, and the solid line corresponds to
%WJ    the covariance-weighted mean $\bar{\nu}_\mathrm{cov}$. 
    The horizontal lines indicate
    the plain average $\bar{\nu}_\mathrm{plain}$ (dotted),
    the error-weighted mean $\bar{\nu}_\mathrm{err}$ (dashed), and
    the covariance-weighted mean $\bar{\nu}_\mathrm{cov}$ (solid). 
    The shaded areas indicate
    the corresponding one-sigma environments $\bar{\nu}\pm \sigma_\mathrm{corr}$.\label{fig:constant_fit}}
\end{figure}

For the case of the 2D model, simulations were performed at the asymptotic critical
coupling $\beta_c = \frac{1}{2}\ln(1+\sqrt{2}) = 0.440686794\ldots$, using a series
of square lattices of linear size $L = 16$, $24$, $\ldots$, $192$.
%WJ Different estimates of $\nu$ were then extracted from fits of the functional
%WJ forms (\ref{eq:binder_scaling}) and (\ref{eq:logmagnderiv_scaling}) to the maximal
%WJ values of these quantities. 
For this model, and the considered range of system sizes, we do not find corrections
to scaling to be very pronounced, such that high-quality fits can be achieved while
ignoring the terms proportional to $L^{-w}$ in (\ref{eq:binder_scaling}) and
(\ref{eq:logmagnderiv_scaling}) and restricting the range of system sizes to $L \ge
32$. The resulting estimates are collected in the left part of Table
\ref{tab:2D_3D_av}. Table \ref{tab:2D_corr} shows the matrix of correlation
coefficients $\rho_{ij} = \Gamma_{ij}/\sigma_i\sigma_j$ for these estimates as
computed from the jackknife approach (\ref{eq:COV}). It maybe does not come
unexpected that all of the estimates for $\nu$, resulting from structurally similar
observables in the magnetic sector, are highly correlated with $\rho_{ij} \gtrsim
0.8$. One might naturally wonder, then, if it is indeed worthwhile to consider all of
these different estimators instead of, say, the single most precise one. The
different averages discussed above are listed in the lower part of Table
\ref{tab:2D_3D_av} together with the error estimates $\sigma_\mathrm{uncorr}$
neglecting correlations and $\sigma_\mathrm{corr}$ taking them into account. For the
plain average as well as the error-weighted mean it is apparent that, although
$\sigma_\mathrm{uncorr}$ seems to indicate smaller fluctuations than for any single
estimate, using the proper error $\sigma_\mathrm{corr}$ the situation is reversed and
the uncertainties of some of the single estimates, namely those stemming from the
logarithmic derivatives, are smaller than the true fluctuation of these averages. For
the full covariance-weighted mean, on the other hand, one arrives at
$\bar{\nu}_\mathrm{cov} = 0.9935(78)$, which has clearly smaller fluctuations than
any of the individual estimates. As is apparent from the lower part of Table
\ref{tab:2D_corr}, this improvement is effected through a dramatically different
choice of weights for the individual estimates as compared to the error-weighting or
plain-average schemes. Comparing the standard deviations of the most commonly used
average $\bar{\nu}_\mathrm{err}$ and the new $\bar{\nu}_\mathrm{cov}$, it is striking
that statistical precision is increased by almost a factor of three merely by using
different weights in the average. Against our usual intuition developed from
statistics of uncorrelated events, the average $\bar{\nu}_\mathrm{cov}$ is here found
to be smaller than all individual estimates. This is illustrated in
Fig.~\ref{fig:constant_fit}, where $\bar{\nu}_\mathrm{cov}$ can also be interpreted
as a correlated fit to a constant (see also Ref.~\cite{michael:94}).

\begin{table}[tb]
  \centering
  \caption
  {
    Correlation coefficients $\rho_{ij} = \Gamma_{ij}/\sigma_i\sigma_j$ between
    estimates of the critical exponent $\nu$ of the 2D Ising model extracted from the
    maxima (\ref{eq:binder_scaling}) and
    (\ref{eq:logmagnderiv_scaling}). The lower part of the table shows the weights
    $\alpha_i$ of the individual estimates
    in the plain, error-weighted and covariance-weighted averages, respectively.
  }
  %\footnotesize
  \begin{ruledtabular}
    \begin{tabular}{cddddd}
      & \multicolumn{1}{c}{$\displaystyle \frac{\d\ln\langle |m|\rangle}{\d\beta}$} & \multicolumn{1}{c}{$\displaystyle \frac{\d\ln\langle m^2\rangle}{\d\beta}$} & \multicolumn{1}{c}{$\displaystyle \frac{\d\ln\langle |m|^3\rangle}{\d\beta}$} & \multicolumn{1}{c}{$\displaystyle \frac{\d U_2}{\d\beta}$} & \multicolumn{1}{c}{$\displaystyle \frac{\d U_4}{\d\beta}$} \\ \colrule
      $\displaystyle \frac{\d\ln\langle |m|\rangle}{\d\beta}$ & 1.000 & 0.974 & 0.939 & 0.920 & 0.897 \\
      $\displaystyle \frac{\d\ln\langle m^2\rangle}{\d\beta}$ & 0.974 & 1.000 & 0.991 & 0.817 & 0.869  \\
      $\displaystyle \frac{\d\ln\langle |m|^3\rangle}{\d\beta}$ & 0.939 & 0.991 & 1.000 & 0.743 & 0.820 \\
      $\displaystyle \frac{\d U_2}{\d\beta}$ & 0.920 & 0.817 & 0.743 & 1.000 & 0.860 \\
      $\displaystyle \frac{\d U_4}{\d\beta}$ & 0.897 & 0.869 & 0.820 & 0.860 & 1.000
      \\ \colrule
      $\alpha_{i,\mathrm{plain}}$ & 1.000 & 1.000 & 1.000 & 1.000 & 1.000 \\
      $\alpha_{i,\mathrm{err}}$ & 0.315 & 0.271 & 0.248 & 0.034 & 0.132 \\
      $\alpha_{i,\mathrm{cov}}$ & 5.007 & -2.426 & -0.281 & -0.104 & -1.196 \\
    \end{tabular}
  \end{ruledtabular}
  \label{tab:2D_corr}
\end{table}

Simulations of the 3D Ising model were performed for simple cubic lattices of edge
lengths $L=8$, $12$, $16$, $\ldots$, $128$ at the fixed coupling $\beta = 0.22165459$
reported in a high-precision study as estimate for the transition point
\cite{bloete:99a}. Here, scaling corrections for the logarithmic derivatives of
magnetization moments are sufficiently pronounced to warrant the inclusion of the
$L^{-w}$ correction term of Eq.~(\ref{eq:logmagnderiv_scaling}). For the cumulants,
corrections are so small that, instead, fits of the uncorrected form were used on the
range $L \ge 32$. The resulting estimates of $\nu$ are collected on the right side of
Table \ref{tab:2D_3D_av}. Concerning the various averages, it is again found that
errors are clearly underestimated when neglecting correlations, and for the plain and
error-weighted means the true fluctuations are indeed larger than the errors of the
single most precise estimate. In contrast, the covariance-weighted mean yields
$\bar{\nu}_\mathrm{cov} = 0.6300(17)$, significantly more precise than the single
estimates as well as the averages not taking correlations into account.

Similar considerations apply to the correlations between estimates of {\em
  different\/} exponents. In particular, taking the scaling relations for critical
exponents into account, the magnetic and energetic scaling dimensions might be
estimated from different observables. For instance, the magnetic scaling dimension
can be estimated via the relations $x_\sigma = \beta/\nu$ and $x_\sigma =
d/2-\gamma/2\nu$ from the FSS of the magnetization at its inflection point and the
magnetic susceptibility at its maximum via the relations $\langle |m|
\rangle_\mathrm{inf}(L) = m_0L^{-\beta/\nu}$ and $\chi_\mathrm{max}(L) = \chi_0
L^{\gamma/\nu}$, respectively. Table \ref{tab:2D_xsigma} summarizes the correlation
analysis for $x_\sigma$ in the 2D model, where through the pronounced {\em
  anti-}correlation of the two estimates of $x_\sigma$ the uncorrelated error
$\sigma_\mathrm{uncorr}$ {\em over-}estimates statistical fluctuations, and already
the error-weighted mean is somewhat more precise than either of the two single
estimates. Still, the covariance-weighted mean is even more precise, yielding
$\bar{x}_{\sigma,\mathrm{cov}} = 0.125(1)$, directly at the exact value $x_\sigma =
1/8$. For the 3D model, on the other hand, (only) the correlation analysis reveals
that both estimates of $x_\sigma$ are nearly uncorrelated such that, for this
specific case, the full result approximately coincides with the naive approach
neglecting correlations.

%\enlargethispage{0.5cm}

To summarize, we have seen that substantial cross-correlations exist between
quantities estimated via histogram analyses from time series of Markov chain 
%WJ Monte Carlo 
MC
simulations.  Taking these into account by a straightforward extension of the
common data analysis reveals a generic underestimation of statistical error by the
conventional approach.  On the other hand, it suggests improved estimators with often
substantially reduced statistical fluctuation resulting, for some examples, in a
threefold reduction of statistical error which could otherwise only be achieved with
an about tenfold increase of simulation time with the conventional analysis. While
these effects have been illustrated here for the case of the critical exponents of
the Ising model, very similar behavior is expected for non-universal quantities,
including properties of first-order transitions \cite{janke:03}, and for different
applications, including the problems in soft-matter systems \cite{holm:05}, for
quantum critical points \cite{vojta:03b}, or the extremely costly simulations of
disordered systems \cite{holovatch:07}. These applications, together with the
flexibility in choosing different thermal or 
%WJ finite-size scaling 
FSS
approaches, render
the outlined technique quite generic.

\begin{table}[t]
  \centering
  \caption
  {
    Correlation analysis and averages of estimates of the magnetic scaling dimension $x_\sigma$ of
    the 2D Ising model from the scaling of the magnetization at its inflection
    point and the magnetic susceptibility at its maximum.
  }
  %\footnotesize
  \begin{ruledtabular}
    \begin{tabular}{lldddd}
      & & \multicolumn{2}{c}{fits} & \multicolumn{2}{c}{corr. coeff./weights}\\
      & & \multicolumn{1}{c}{$x_\sigma$} & \multicolumn{1}{c}{$\sigma$} &
      \multicolumn{1}{c}{$\displaystyle \langle |m| \rangle_\mathrm{inf}$} &
      \multicolumn{1}{c}{$\displaystyle \chi_\mathrm{max}$} \\ \colrule
      \multicolumn{2}{c}{$\displaystyle \langle |m| \rangle_\mathrm{inf}$} & 0.1167 & 0.0054 & 1.0000 & -0.6414 \\ 
      \multicolumn{2}{c}{$\displaystyle \chi_\mathrm{max}$} & 0.1271 & 0.0020 & -0.6414 & 1.0000 \\ \colrule
      $\bar{x}_\mathrm{plain}$ & $\sigma_\mathrm{uncorr}$ & 0.1219 & 0.0027 & 1.0000 & 1.0000\\
      & $\sigma_\mathrm{corr}$ & & 0.0021 &  & \\
      $\bar{x}_\mathrm{err}$ & $\sigma_\mathrm{uncorr}$ & 0.1261 & 0.0016 & 0.0944 & 0.9056\\
      & $\sigma_\mathrm{corr}$ & & 0.0013 &  & \\
      $\bar{x}_\mathrm{cov}$ & $\sigma_\mathrm{corr}$ & 0.1250 & 0.0010 &
      0.2050 & 0.7950\\ \colrule
      \multicolumn{2}{l}{reference value} & 0.125 & & & \\
    \end{tabular}
  \end{ruledtabular}
  \label{tab:2D_xsigma}
\end{table}

M.W.\ acknowledges support by the DFG through the Emmy Noether Programme under
contract No.\ WE4425/1-1.

%\bibliography{general}

\end{document}